\documentclass{aa}
\bibpunct{(}{)}{;}{a}{}{,} 
\usepackage{natbib}
\usepackage{amsmath}
\usepackage{physics}
\usepackage{graphicx}
\usepackage{txfonts}

\begin{document}

\title{The role of the general relativity on icy body reservoirs \\
       under the effects of an inner eccentric Jupiter
      }

   \author{M. Zanardi\inst{1,2}\thanks{mzanardi@fcaglp.unlp.edu.ar},
           G. C. de El\'\i a\inst{1,2},
           R. P. Di Sisto\inst{1,2},
           \and
           S. Naoz\inst{3,4}
          }

   \offprints{M. Zanardi
             }

   \institute{Instituto de Astrof\'{\i}sica de La Plata, CCT La Plata-CONICET-UNLP, Paseo del Bosque S/N (1900), La Plata, Argentina
     \and Facultad de Ciencias Astron\'omicas y Geof\'\i sicas, Universidad Nacional de La Plata, Paseo del Bosque S/N (1900), La Plata\\
     Argentina
   \and Department of Physics and Astronomy, University of California, Los Angeles, CA 90095, USA
   \and Mani L. Bhaumik Institute for Theoretical Physics, Department of Physics and Astronomy, UCLA, Los Angeles, CA 90095, USA
                }

   \date{Received / Accepted}

\abstract
{
Recent studies have analyzed the dynamical evolution of outer small body populations under the effects of an eccentric inner massive perturber, which result from a planetary scattering event. These investigations suggest that such outer reservoirs are composed of particles on prograde and retrograde orbits, as well as particles whose orbit flips from prograde to retrograde and back again showing a coupling between the inclination $i$ and the ascending node longitude $\Omega$ (Type-F particles)
}
{
We analyze the role of the General Relativity (GR) on the dynamics of outer particles under the influence of an inner eccentric Jupiter-mass planet produced by a planetary scattering event. In particular, we are interested to study how the GR affects the dynamical evolution of the outer Type-F particles, which experience an eccentric Lidov-Kozai mechanism.
}
{
To do this, we carry out N-body simulations with and without GR effects. Such a detailed comparative analysis will allow us to strengthen our understanding concerning the GR and eccentric Lidov-Kozai combined effects on the dynamical evolution of outer particles.  
}
{
When the GR is included, the extreme values of $\Omega$ are obtained for retrograde inclinations, while the minimum and maximum inclinations allowed for Type-F particles increase in comparison with that derived without GR effects. According to this, if the GR is included in the simulations, the range of prograde (retrograde) inclinations of the libration region is reduced (increased) respect to that obtained in absence of GR. We find two new class of particles when the GR effects are included in the simulations. On the one hand, particles whose orbital plane flips from prograde to retrograde and back again without experiencing a coupling between $i$ and $\Omega$. On the other hand, retrograde particles that show a strong coupling between $i$ and $\Omega$. We infer that the GR may significantly modify the dynamical properties of the outer reservoirs that evolve under the effects of an eccentric inner perturber.
}
{}
%


\authorrunning{M. Zanardi, et al.
               }
\titlerunning{
GR and eccentric giant planet effects on outer reservoirs of minor planets
             }

\maketitle
\section{Introduction}

The secular dynamics of a system composed of an outer body orbiting around an inner binary results to be complex. Most of the studies carried out to date analyzed the effects of a far away massive perturber on the inner binary. In particular, \citet{Lidov1962} and \citet{Kozai1962} developed pioneering works aimed at analyzing the secular perturbations of an inner test particle produced by an outer massive perturber on a circular orbit. In this scenario of work, the authors showed that the test particle conserves the vertical angular momentum, and thus, its orbit oscillates periodically, exchanging eccentricity for inclination. An interesting result derived by \citet{Kozai1962} indicates that if the particle's inclination is between 39.2$^{\circ}$ and 140.8$^{\circ}$, its pericentre argument shows librations coupled with the inclination and the eccentricity. In this libration regime, \citet{Kozai1962} showed that particles of very low eccentricity can not remain on a circular orbit along its evolution. In particular, if the particle's inclination is close to 90$^{\circ}$, its maximum eccentricity is nearly unity. This dynamical behavior is so-called {\it standard Lidov-Kozai mechanism}.

During the last years, the {\it eccentric Lidov-Kozai mechanism} for an inner test particle has been the focus of several investigations. In particular, \citet{Lithwick2011} and \citet{Katz2011} studied the dynamical behavior of a massless particle that orbits a star under the effects of an outer perturber up to the octupole level of the secular approximation. Such works showed that if the planet has an eccentric orbit, the vertical angular momentum is no longer conserved and the eccentric Lidov-Kozai mechanism can lead to a peculiar evolution on the test particle, whose orbit can flip reaching prograde and retrograde inclinations together with high eccentricities. Many others authors have developed important progresses concerning the effects of an outer and eccentric perturber on the evolution of an inner test particle. In fact, \citet{Libert2012} analyzed the dynamical evolution of an inner body with high inclination in the presence of an eccentric and outer perturber. Moreover, \citet{Li2014} developed an study about chaos in the eccentric Lidov-Kozai mechanism in the inner test particle limit. In addition, \citet{Antognini2015} derived expressions for the timescales of Lidov-Kozai oscillations up to the quadrupole and octupole level in the test particle case. A detailed description about the eccentric Lidov-Kozai mechanism and its applications for an inner test particle can be found in \citet{Naoz2016}.

The secular perturbations of an outer test particle that evolves under the effects of an inner and eccentric perturber have been also studied in the literature for several authors. First, \citet{Ziglin1975} analyzed the secular evolution of a far away planet in a binary-star system in the context of the restricted elliptical three-body problem. From a doubly averaged disturbing function, the author found that the dynamical behavior of the outer test particle is strongly dependent on the eccentricity of the inner binary. In fact, if the binary's orbit is circular, particle's node always precesses, while if the binary's orbit is elliptic, nodal librations take place. Moreover, \citet{Ziglin1975} derived an expression for the width of the libration zone observing that the greater the eccentricity of the inner binary, the bigger the libration region. Then, \citet{Thomas1996} studied the secular evolution of the minor planets of the outer Solar System under the perturbations of the four giant planets on coplanar and circular orbits. They found that the Kozai resonance fails to excite the eccentricities from the quasi-circular orbits even with very large inclinations. However, the resonance modifies the dynamical evolution of the very eccentric orbits, such as those associated to the large period comets. Later, \citet{Farago2010} studied the case of a far away body orbiting an inner binary up to the quadrupole level of the secular approximation in the context of the three-body problem. For the particular case where the outer body is a massless test particle, \citet{Farago2010} obtained results consistent with those derived by \citet{Ziglin1975}. Recently, \citet{Naoz2017} and \citet{Zanardi2017} (hereafter {\it Paper 1}) carried out two complementary works aimed at analyzing the dynamical evolution of outer test particles under the effects of an eccentric inner massive perturber. In fact, \citet{Naoz2017} studied the secular evolution of a far away test particle orbiting an inner massive binary up to the octupole level of approximation. From this analysis, the angular momentum of the outer test particle is conserved at the quadrupole level, which indicates that its eccentricity remains constant. However, \citet{Naoz2017} showed that secular interactions up to the octupole level lead to variations of the test particle's eccentricity and introduces higher-level resonances, which may produce chaos. In this context, \citet{Zanardi2017} made use of N-body simulations and analyzed the dynamical evolution of outer test particles in presence of an eccentric inner Jupiter-mass planet resulting from a planet-planet scattering event around a 0.5 M$_{\odot}$ star. The results derived by these authors suggest that planetary scattering scenarios lead to an efficient production of outer small body reservoirs composed of particles with prograde and retrograde inclinations, and other ones whose orbit flips from prograde to retrograde and back again in their evolution. The dynamical features of these flipping particles, which show a strong correlation between the inclination $i$ and the longitude of ascending node $\Omega$, were supported by the analytical results derived by \citet{Naoz2017}. Such particles, whose orbital plane flips showing a coupling between $i$ and $\Omega$, were called {\it Type F-Particles} by \citet{Zanardi2017}.

One of the effects that can modify the dynamical properties of a given system derived from the eccentric Lidov-Kozai mechanism is the General Relativity (GR) \citep{Ford2000_1,Ford2000_2,Naoz2013,Sekhar2017}. For the particular case of an outer test particle orbiting an inner binary, \citet{Naoz2017} suggested that the inclination excitation can be suppressed (produced) for systems with GR precession faster (slower) than the quadrupole precession. According to this, the GR effects may significantly modify the dynamical structure of outer small body reservoirs that evolve under the influence of inner eccentric perturber.

The main goal of the present research is to analyze the role of the GR on the dynamics of outer particles under the influence of an inner eccentric Jupiter-mass planet. In particular, we are interested to study how the GR affects the dynamical evolution of the outer Type-F particles, which experiment an eccentric Lidov-Kozai mechanism. To do this, we include the GR forces in the N-body code and rerun those simulations of the Paper 1 that produce systems where at least one Type-F particle exists in the resulting small body population, which evolve under the influence of an inner eccentric Jupiter-mass planet. A detailed comparative analysis between the N-body simulations without GR and the GR-included simulations carried out in the present research will allow us to strengthen our understanding concerning the GR and eccentric Lidov-Kozai combined effects on the dynamical evolution of outer particles. The present paper is structured as follows. In Section 2, we present the scenario of our simulations and the numerical code used for carrying out the N-body integrations. A detailed analysis of the results about the dynamical properties of the small body reservoirs resulting from our numerical simulations is shown in Section 3. Finally, Section 4 describes the discussions and conclusions of our study.

\begin{figure}
\centering
\includegraphics[angle=270, width=0.48 \textwidth]{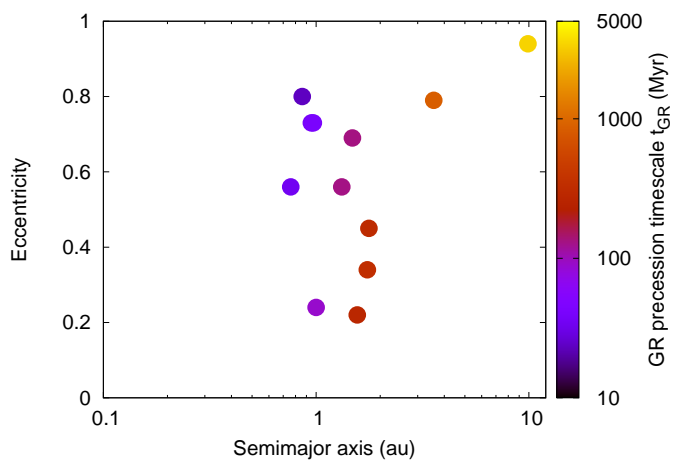}
\caption{Orbital parameters $e$ vs. $a$ of the surviving planets in each of the 12 ``Type-F simulations''. The color palette represents the timescale of GR (Eq.~\ref{eq:GR-timescale}), which leads to the precession of the planet's pericentre argument. The orbital elements illustrated here correspond to those adopted for each surviving planet immediately after the dynamical instability event simulated in Paper 1.   
}
\label{fig:fig1}
\end{figure}

\begin{figure*}
\centering
\includegraphics[angle=270, width=0.98 \textwidth]{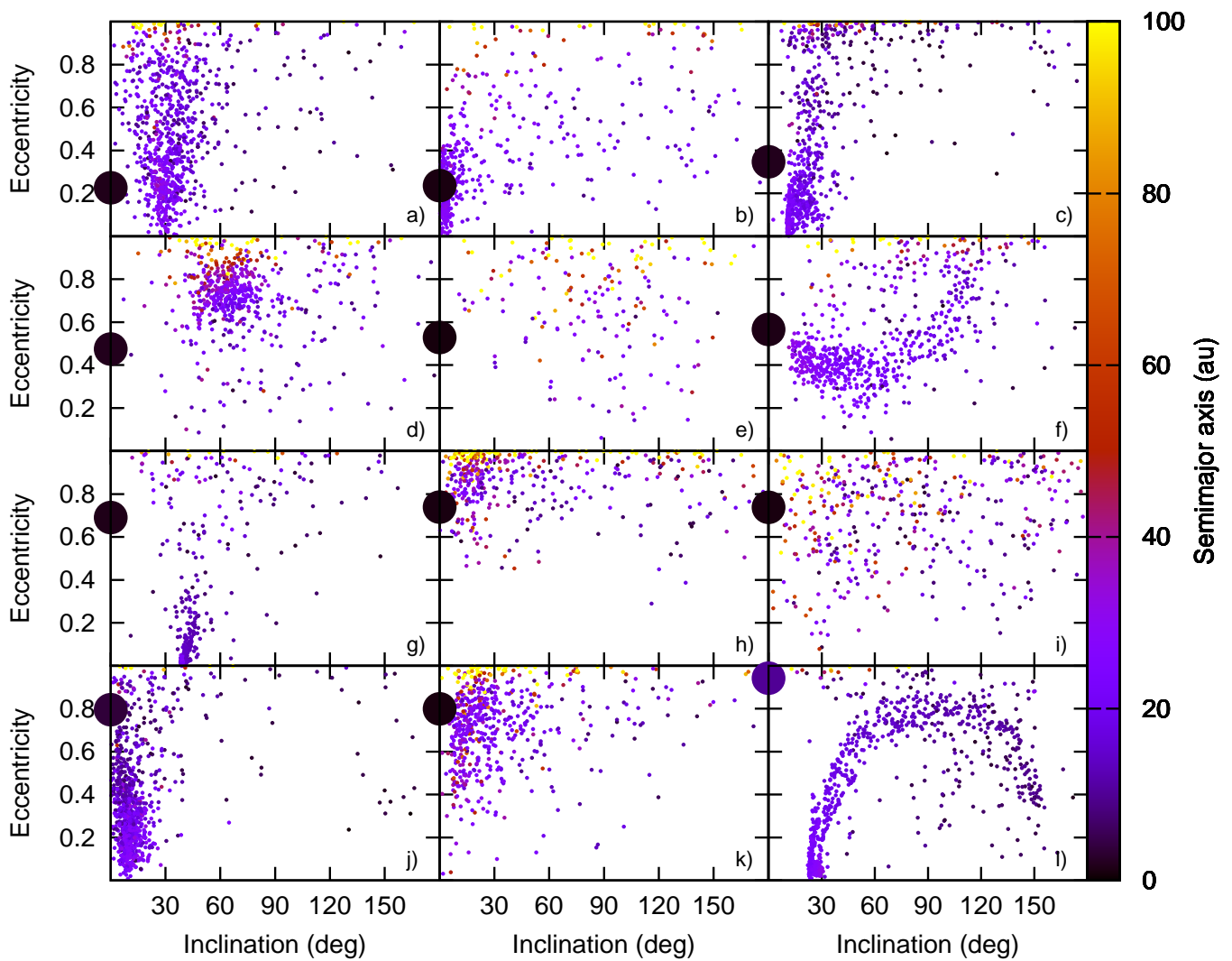}
\caption{Orbital elements $i$ vs. $e$ of all particles (small circles) and the surviving Jupiter-mass planet (big circle) immediately after the instability event for each of the 12 ``Type-F simulations'' of the Paper 1. The color palette represents the semimajor axis of all particles and of the giant planet of each system immediately after the instability event.
}
\label{fig:fig2}
\end{figure*}

\section{N-body simulations: scenarios of work and numerical code}

In Paper 1, \citet{Zanardi2017} analyzed the evolution of systems initially composed of three Jupiter-mass planets located close to their dynamical instability limit together with an cold outer disk of test particles around a 0.5 M$_{\odot}$ star, which undergo strong planetary scattering events. In particular, that work based on the analysis of 21 N-body simulations, in which a single Jupiter-mass planet survives in the system after the dynamical instability event. Thus, the study developed by \citet{Zanardi2017} offers a detailed description about the main dynamical features of outer test particles evolving under the influence of an eccentric inner perturber. According to that shown in Paper 1, the resulting outer reservoirs are composed only of particles on prograde orbits (hereafter {\it Type-P particles}) in 7 of 21 simulations, while in 2 of 21, Type-P particles coexist with particles on retrograde orbits (hereafter {\it Type-R particles}). The most important result derived in Paper 1 is the existence of Type-F particles with very peculiar dynamical properties, which are present in 12 of those 21 N-body simulations with a single surviving Jupiter-mass planet. As we have already mentioned, the Type-F particles have orbits that flip from prograde to retrograde and back again showing a coupling between the inclination $i$ and the ascending node longitude $\Omega$. In general terms, the Type-F particles coexist with Type-P and -R particles in the outer small body reservoirs that result of those 12 N-body simulations, which emerge as very auspicious scenarios to analyze the dynamical features of the different classes of particles. The present work is based on the 12 aforementioned simulations, which are called ``Type-F simulations''.

Figure~\ref{fig:fig1} illustrates the distribution of each surviving Jupiter-mass planet in the orbital element space eccentricity $e$ vs. semimajor axis $a$, which are associated to the 12 ``Type-F simulations''. Such orbital elements correspond to those adopted for each surviving planet immediately after the dynamical instability event simulated in Paper 1. These parameters are referenced to the barycentre and invariant plane of the system, whose x-axis coincides with the pericentre of the planet. The semimajor axis $a$ and eccentricity $e$ of such planets range from 0.76 au to 10 au and 0.22 to 0.94, respectively. In such a figure, the color palette represents the orbit GR precession timescale in a first order post-Newtonian approximation, which can be estimated as
\begin{eqnarray}
\centering
t_{\text{GR}} = 2 \pi \frac{a^{5/2} c^{2} (1-e^{2})}{3 k^{3} (m_{\star} + m)^{3/2}},
\label{eq:GR-timescale}  
\end{eqnarray}   
\citep{Einstein1916} where $a$, $e$, and $m$ represent the semimajor axis, the eccentricity, and the mass of the planet, respectively, $m_{\star}$ is the mass of the star, $c$ the speed of light, and $k^{2}$ the gravitational constant. As the reader can see, the orbit GR precession takes place on timescales that range from 23.4 Myr to 3400 Myr for the planets represented in Fig.~\ref{fig:fig1}. As it is shown in \citet{Naoz2017}, the GR precession of the inner planet's pericentre argument translates to a precession of the outer particle's ascending node longitude. Taking into account the strong correlation between the orbital inclination and the ascending node longitude of the Type-F particles \citep{Zanardi2017,Naoz2017}, the GR may play an important role in the dynamical properties of such particles as well as in the global structure of the outer small body reservoirs that evolve under the influence of an inner eccentric giant planet.

To study this, we carry out two new sets of N-body simulations, which are based on the 12 ``Type-F simulations''. The first set of simulations only includes gravitational forces in the evolution of each of the 12 aforementioned systems (hereafter {\it No-GR simulations}), while the second one incorporates GR effects in the integration of each of such systems (hereafter {\it GR simulations}). We remark that all N-body simulations are carried out using the MERCURY code \citep{Chambers1999}. In particular, we make use of the Bulirsch-Stoer algorithm with an accuracy parameter of 10$^{-12}$. To include the GR effects, we modify the MERCURY code including an additional acceleration from the GR corrections derived by \citet{Anderson1975}, which is given by
\begin{eqnarray}
\centering
  \Delta{{\ddot{\bf r}}} = \frac{k^{2}m_{\star}}{c^2 r^3} \Bigg \{ \left( \frac{4k^{2}m_{\star}}{r} - {\bf v}\cdot {\bf v} \right) {\bf r} + 4\left({\bf r}\cdot {\bf v} \right){\bf v} \Bigg \},
\label{eq:GR-Anderson}
\end{eqnarray}
where $m_{\star}$ represents the mass of the star, $k^{2}$ the gravitational constant, $c$ the speed of light, {\bf r} and {\bf v} the astrocentric position and velocity vectors, respectively, and $r = \mid {\bf r} \mid$. It is worth noting that this is a first order post-Newtonian approximation. The hamiltonian and its connection to the three-body problem are shown in \citet{Naoz2013}.

The No-GR and GR simulations of the present work are developed assuming as initial conditions the orbital parameters of the systems immediately after the instability event when a single planet survives in the system (hereafter {\it post IE orbital parameters}), for each of the 12 ``Type-F simulations'' of the Paper 1. Figure~\ref{fig:fig2} shows the post IE inclination $i$ and eccentricity $e$ of all particles and the eccentric inner perturber of the 12 ``Type-F simulations'', where the color palette represents the post IE semimajor axis $a$ of the particles and of the surviving giant planet. It is worth noting that the orbital elements of all bodies illustrated in Fig.~\ref{fig:fig2} are referenced to the barycentre and invariant plane of the system, where x-axis is oriented to the surviving planet's pericentre.

All N-body simulations of the present research are integrated for a time span of 2$t_{\text{GR}}$ in order to explore the role of the GR in the dynamical evolution of the test particles.
\section{Results}
\label{sec:result}

Here, we describe the results of our N-body simulations with the aim of analyzing the role of the GR on the dynamical properties of the outer test particles that evolve under the effects of an eccentric inner Jupiter-mass planet. Hereafter, the subscripts 1 and 2 will be used to refer to the mass and the orbital parameters of the inner planet and the outer test particles, respectively. Moreover, we remark that the orbital parameters of outer test particles are referenced to the barycentre and invariant plane of the system, where x-axis coincides with the planet's pericentre. It is worth noting that, when the GR is included, the inner planet's pericentre argument precesses for which the ascending node longitude of the test particles will be referenced to a rotating system.

\begin{figure}
\centering
\includegraphics[angle=270, width=0.48 \textwidth]{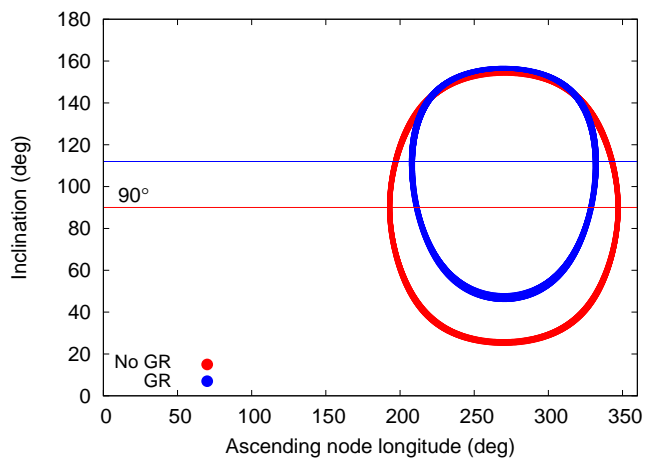}
\caption{Coupling between the inclination $i_2$ and the ascending node longitude $\Omega_2$ of a given Type-F particle, which result from the No-GR (red curve) and GR (blue curve) simulations. The initial orbital elements of this particle are $a_{2}$ = 20.72 au, $e_{2}$ = 0.62, $i_{2}$ = 144.84$\degr$, $\omega_{2}$ = 42.15$\degr$ and $\Omega_{2}$ = 222.30$\degr$. The inner planet of such a system has a semimajor axis $a_{1} =$ 0.95 au and an eccentricity $e_{1} =$ 0.74. The red and blue horizontal lines represent the inclination values for which the ascending node longitude reaches extreme values for No-GR and GR simulations, respectively. 
}
\label{fig:fig3}
\end{figure}

\subsection{Role of the GR on the dynamical evolution of Type-F particles} 

Figure~\ref{fig:fig3} shows trajectories of evolution of a given Type-F particle in an inclination $i_2$ vs. ascending node longitude $\Omega_2$ plane, which resulting from the No-GR (red curve) and GR (blue curve) simulations. 

According to \citet{Naoz2017}, the time evolution of the inclination $i_{2}$ and the ascending node longitude $\Omega_{2}$ of an outer test particle up to the quadrupole level of secular approximation without GR is given by
\begin{eqnarray}
  \left(\frac{di_{2}}{dt}\right)_{\text{quad}} &=& \frac{m_{1}m_{\star}}{(m_{1} + m_{\star})^{2}} \frac{2\pi}{P_{2}}  \left( \frac{a_{1}}{a_{2}} \right)^2    \nonumber \\ 
   &\times& \frac{15e^{2}_{1}\sin i_{2}\sin 2\Omega_{2}}{8(1 - e_{2}^{2})^2},   
   \label{eq:eq3} \\
  \left(\frac{d\Omega_{2}}{dt}\right)_{\text{quad}} &=& - \frac{m_{1}m_{\star}}{(m_{1} + m_{\star})^{2}} \frac{2\pi}{P_{2}}  \left( \frac{a_{1}}{a_{2}} \right)^2 \nonumber \\ 
  &\times& \frac{3\cos i_{2}(2 + 3e_{1}^{2} - 5e_{1}^{2}\cos2\Omega_{2})}{8(1 - e_{2}^{2})^2},
\label{eq:eq4}  
\end{eqnarray}
where $m_{1}$, $a_{1}$, and $e_{1}$ are the mass, semimajor axis, and eccentricity of the inner planet, respectively, $a_{2}$, $e_{2}$, and $P_{2}$ the semimajor axis, eccentricity, and orbital period of the outer test particle, and $m_{\star}$ the mass of the central star. On the one hand, the minimum and maximum inclinations for libration trajectories are obtained for values of the ascending node longitude $\Omega_{2}$ of 90$^{\circ}$ or 270$^{\circ}$. On the other hand, the extreme values of the ascending node longitude $\Omega_{2}$ for libration trajectories are associated to an inclination $i_{2} =$ 90$^{\circ}$. These correlations between $i_{2}$ and $\Omega_{2}$ in absence of GR can be seen in the red curve represented in the Fig.~\ref{fig:fig3}.

The inclusion of GR modifies the orbital properties of a Type-F particle concerning the evolution of its inclination $i_{2}$ and ascending node longitude $\Omega_{2}$. On the one hand, the time evolution of the inclination $i_{2}$ of an outer test particle for the quadrupole-level of approximation with GR does not show changes in comparison with Eq.~\ref{eq:eq3}. Thus, the minimum and maximum inclinations for libration trajectories are also obtained for values of the ascending node longitude $\Omega_{2}$ of 90$^{\circ}$ or 270$^{\circ}$. On the other hand, the time evolution of the ascending node longitude $\Omega_{2}$ of an outer test particle for the quadrupole-level of approximation significantly changes with the inclusion of GR. In fact, \citet{Naoz2017} showed that the GR precession of the inner planet's pericentre argument $\omega_1$ translates to a precession of the outer particle's ascending node longitude $\Omega_2$. Thus, working in the rotating frame of the inner planet's orbit, the time evolution of the ascending node longitude $\Omega_{2}$ due to the secular evolution up to the quadrupole level of approximation together with the precession of the ascending node longitude induced by GR effects is given by
\begin{eqnarray}
  \frac{d\Omega_{2}}{dt} = \left(\frac{d\Omega_{2}}{dt}\right)_{\text{quad}} + \left(\frac{d\Omega_{2}}{dt}\right)_{\text{GR}},
  \label{eq:eq5}
\end{eqnarray}
where $\left(\frac{d\Omega_{2}}{dt}\right)_{\text{quad}}$ is given by Eq.~\ref{eq:eq4}, and $\left(\frac{d\Omega_{2}}{dt}\right)_{\text{GR}}$ is expressed by
\begin{eqnarray}
 \left(\frac{d\Omega_{2}}{dt}\right)_{\text{GR}}  = -3 k^{3} \frac{(m_{1} + m_{\star})^{3/2}}{a_{1}^{5/2} c^{2} (1 - e_{1}^{2})},
 \label{eq:evo-node2-GR}
\end{eqnarray}
\citet{Naoz2017}. Notice that the minus sign of this equation indicates that the ascending node longitude's precession of the outer test particle is in the opposite direction to the pericentre argument's precession of the inner planet, which is caused by GR effects. From this, we can find the minimum and maximum ascending node longitude for libration trajectories by setting $\dot \Omega_{2}$  $=$ 0 in the Eq.~\ref{eq:eq5}. Thus, the values of the inclination $i_{2}$ that satisfy this condition are given by the following expression 
\begin{eqnarray}
 i^{*}_{2} =  \arccos \left(A \frac{a_{2}^{7/2}(1 - e_{2}^{2})^{2}}{a_{1}^{9/2}(1 - e_{1}^{2})(2 + 3e_{1}^{2} - 5e_{1}^{2} \cos2\Omega_{2})} \right),
 \label{eq:imin}
\end{eqnarray}
where $A$ is a constant that depends on the mass of the inner planet $m_{1}$ and the mass of the star $m_{\star}$, and it is given by  
\begin{eqnarray}
  A = -\frac{8 k^{2}(m_{1} + m_{\star})^{3}}{c^{2} m_{1}m_{\star}},
  \label{eq:constante-A}
\end{eqnarray}
being $c$ the speed of light, and $k^{2}$ the gravitational constant. Since $A <$ 0, the extreme values of the ascending node longitude $\Omega_{2}$ are obtained for values of the inclination $i_{2}$ higher than 90$^{\circ}$ when the GR effects are included. This shift of the inclination from 90$^{\circ}$ towards retrograde values can be seen in the blue curve represented in the Fig.~\ref{fig:fig3}, which illustrates the evolution of a given outer test particle including GR effects. From Eq.~\ref{eq:imin}, it is worth noting that the values of the inclination $i_{2}$ that lead to extreme values of the ascending node longitude $\Omega_{2}$ depend on the semimajor axis $a_{1}$ and the eccentricity $e_{1}$ of the inner planet, as well as on the semimajor axis $a_{2}$, the eccentricity $e_{2}$, and the ascending node longitude $\Omega_{2}$ of the outer test particle. Thus, for given values of $a_{1}$ and $e_{1}$, the larger the semimajor axis $a_{2}$ and the smaller the eccentricity $e_{2}$, the higher the inclination $i_{2}$ for which the ascending node longitude $\Omega_{2}$ reaches extreme values. Moreover, the dependence of Eq.~\ref{eq:imin} on $\Omega_{2}$ indicates that the value of the inclination $i_{2}$ associated to extreme values of the ascending node longitude $\Omega_{2}$ changes while the particle's orbit flips.

Another important point of our analysis is related with the minimum and maximum inclination that can reach one Type-F particle under the influence of an inner planet with a given eccentricity. In absence of GR, \citet{Naoz2017} showed that the secular Hamiltonian up to the quadrupole level of approximation is expressed by
\begin{eqnarray}
  f_{\text{quad}} = \frac{(2 + 3e^{2}_{1})(3\cos^{2}i_{2} - 1) + 15e^{2}_{1}(1 - \cos^{2}i_{2})\cos2\Omega_2}{(1 - e^{2}_{2})^{3/2}}.
  \label{eq:fquad}
\end{eqnarray}       
This $f_{\text{quad}}$ is an energy function that can be used to determine the extreme values of the inclination $i_{2}$ for libration trajectories, which are obtained for values of the ascending node longitude $\Omega_{2}$ of 90$^{\circ}$ or 270$^{\circ}$. Thus, considering the conservation of energy on the separatrix between the minimum value of the ascending node longitude ($\Omega_{2} =$ 0$^{\circ}$, $i_{2} =$ 90$^{\circ}$) and the extreme values of the inclination ($\Omega_{2} =$ 90$^{\circ}$, $i_{2} = i^{\text{e}}_{2}$), it is possible to obtain that
\begin{eqnarray}
 i^{\text{e}}_{2} = \arccos \left(\sqrt{\frac{5e^{2}_{1}}{(1 + 4e^{2}_{1})}} \right).
\label{eq:imin-sinGR}
\end{eqnarray}
Thus, in absence of GR, the extreme values of the inclination $i_{2}$ on libration trajectories only depend on the eccentricity $e_{1}$ of the inner planet \citep{Ziglin1975,Farago2010,Naoz2017,Zanardi2017}.

\begin{figure}
\centering
\includegraphics[angle=270, width=0.48 \textwidth]{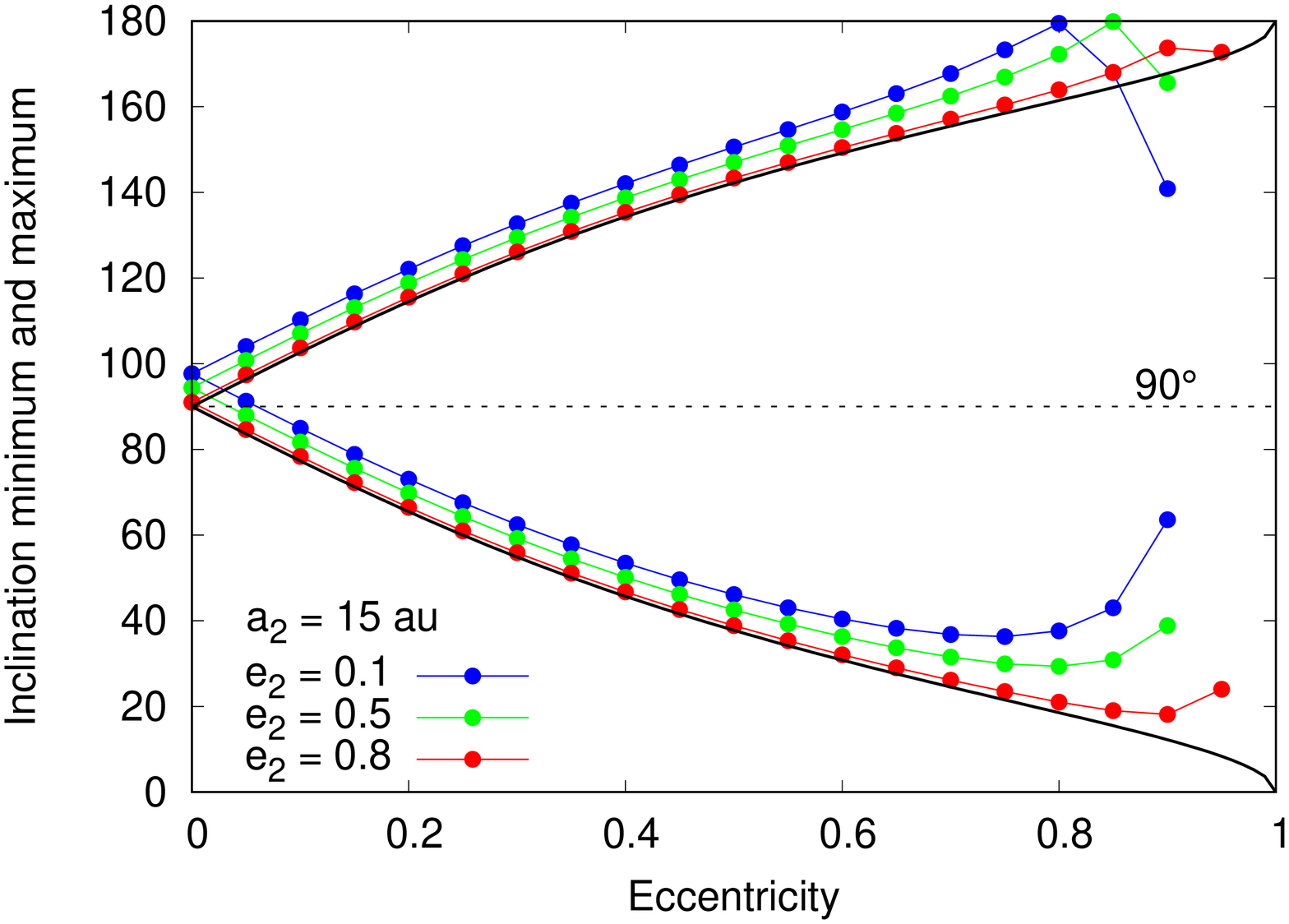}
\includegraphics[angle=270, width=0.48 \textwidth]{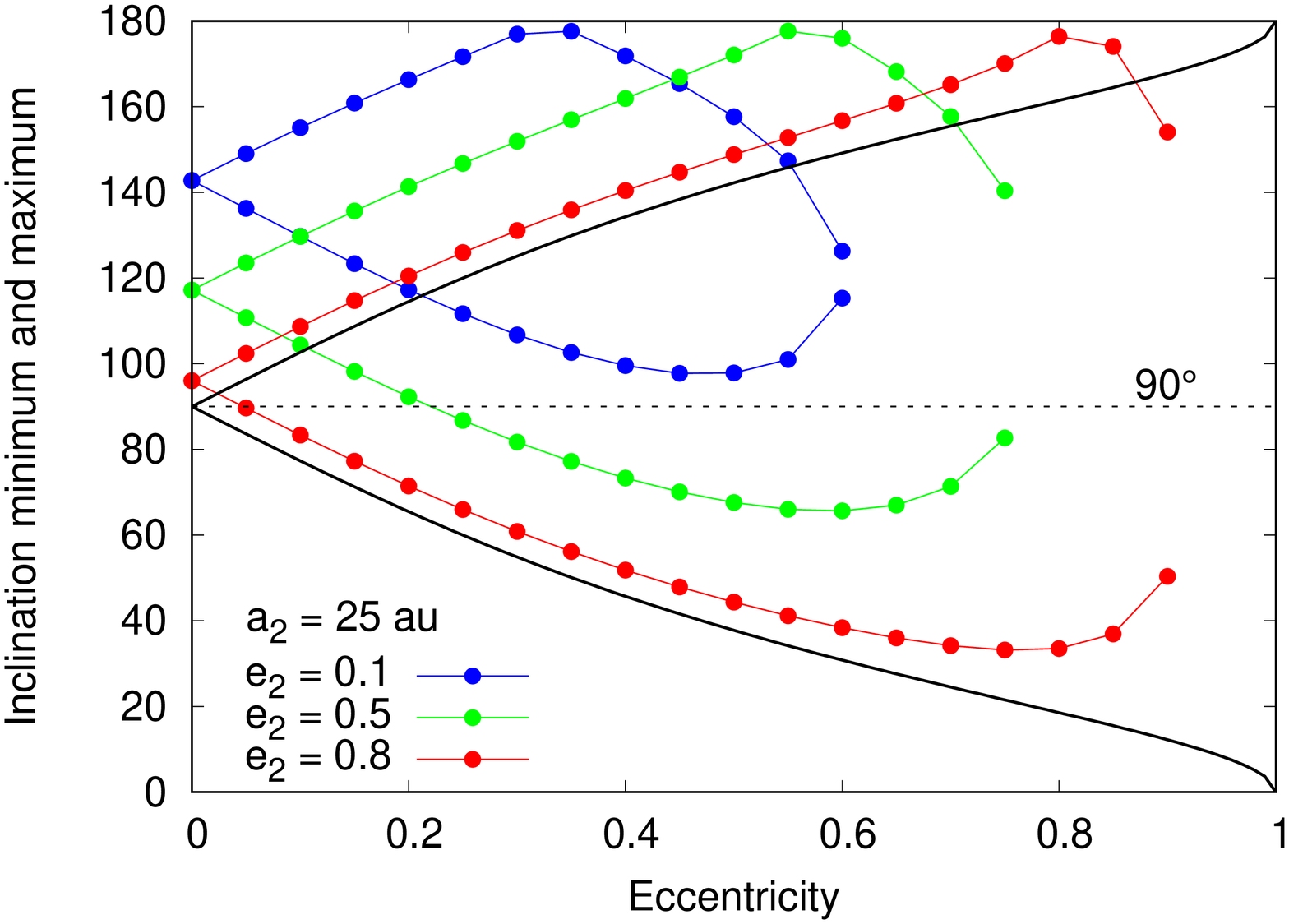}
\caption{The minimum and maximum inclination allowed for libration trajectories as a function of the planet's eccentricity. The black curves represent the correlation in absence of GR. The blue, green, and red curves in both panels show the extreme values of the inclination for eccentricity $e_{2}$ of 0.1, 0.5 and 0.8, respectively, when the GR effects are included. Moreover, we adopt a particle's semimajor axis $a_{2}$ of 15 au and 25 au in the top and bottom panels, respectively. Finally, we consider a semimajor axis $a_{1} =$ 1 au for the planet in both panels.
}
\label{fig:fig4}
\end{figure}

\begin{figure*}
\centering
\includegraphics[angle=270, width=0.98 \textwidth]{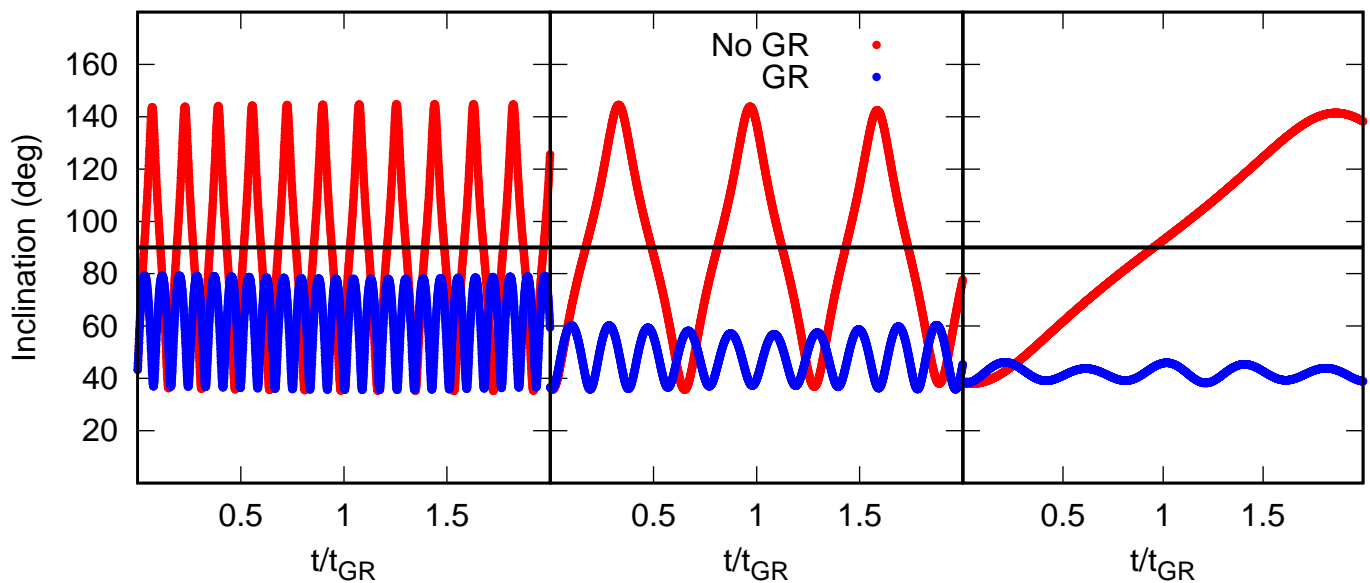}
\caption{Evolution in time of the inclination (normalized to $t_{\text{GR}}$) of three different particles of the same system, which were Type-F particles in absence of GR and then were suppressed by GR effects. The blue and red curves represent the temporal evolution of the particles's inclination with and without GR effects. The semimajor axis and eccentricity of the inner planet are $a_{1}=$ 1.32 au, $e_{1}=$ 0.56, and the orbit GR precession timescale is $t_{\text{GR}} =$ 130 Myr. Left panel: the initial orbital parameters of the particle are $a_{2}=$ 20.74 au, $e_{2}=$ 0.42, $i_{2}=$ 43.30$\degr$, $\omega_{2}=$ 108.35$\degr$ and $\Omega_{2}=$ 233.28$\degr$. Middle panel: the initial orbital parameters of this particle are $a_{2}=$ 44.25 au, $e_{2}=$ 0.724, $i_{2}=$ 36.65$\degr$, $\omega_{2}=$ 290.77$\degr$ and $\Omega_{2}=$ 284.67$\degr$. Right panel: the initial orbital parameters of the outer test particle are $a_{2}=$ 160. au, $e_{2}=$ 0.93, $i_{2}=$ 38.39$\degr$, $\omega_{2}=$ 287.95$\degr$ and $\Omega_{2}=$ 279.02$\degr$.
}
\label{fig:fig3.2.1}
\end{figure*}

According to the equations of motion up to the quadrupole level of approximation proposed by \citet{Naoz2017},
\begin{eqnarray}
 \left(\frac{d\Omega_{2}}{dt}\right)_{\text{quad}}  = \frac {k}{16(1-e^{2}_{2})^{1/2}} \frac{m_{1}m_{\star}}{(m_{1} + m_{\star})^{3/2}} \frac{a^{2}_{1}}{a^{7/2}_{2}} \frac{1}{\sin i_{2}} \frac{\partial f_{\text{quad}}}{\partial i_{2}},
\label{eq:eq11}
\end{eqnarray}
where $f_{\text{quad}}$ is given by the Eq.~\ref{eq:fquad}. When the GR effects are included in the analysis and working in the rotating frame of the inner planet's orbit, an integral of motion $f$ appears which adopts the following expression
\begin{eqnarray}
f = f_{\text{quad}} + 48 \cos i_{2} \frac{k^2}{c^2}\frac{(m_{1}+m_{\star})^3}{m_{1}m_{\star}} \frac{(1-e^{2}_{2})^{1/2}}{(1-e^{2}_{1})} \frac{a^{7/2}_{2}}{a^{9/2}_{1}}. 
\label{eq:f-energia}
\end{eqnarray}  
As the reader can see, if the integral of motion $f$ given by Eq.~\ref{eq:f-energia} replaces to $f_{\text{quad}}$ in Eq.~\ref{eq:eq11}, the expression of $\left(\frac{d\Omega_{2}}{dt}\right)$ given by Eq.~\ref{eq:eq5} is obtained. A general detailed study about the secular and hierarchical three-body problem including post-Newtonian corrections can been found in \citet{Naoz2013}.

\begin{figure*}
\centering
\includegraphics[angle=270, width=0.98\textwidth]{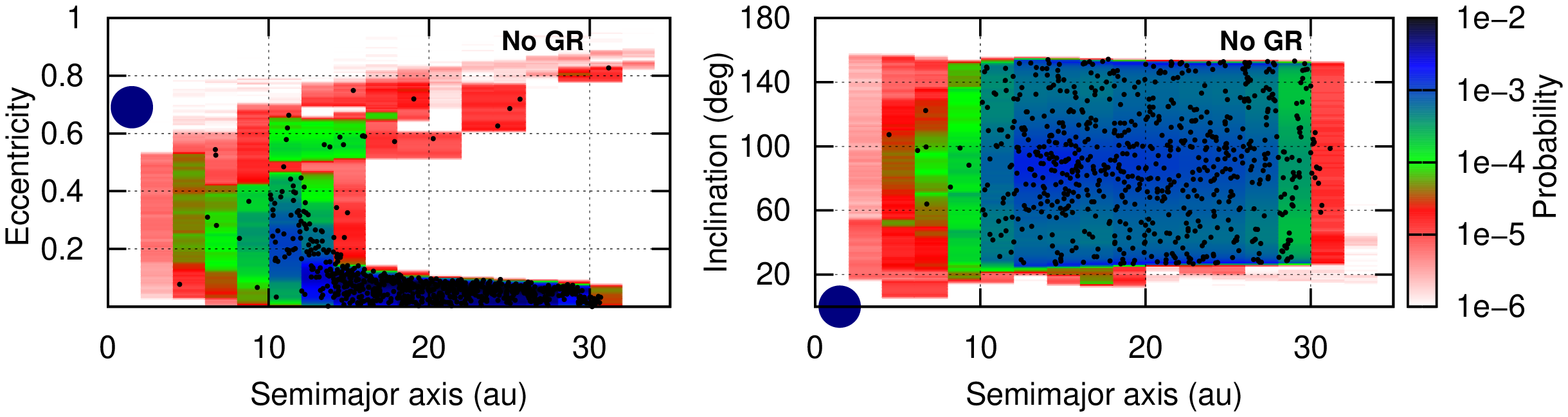}\\
\includegraphics[angle=270, width=0.98\textwidth]{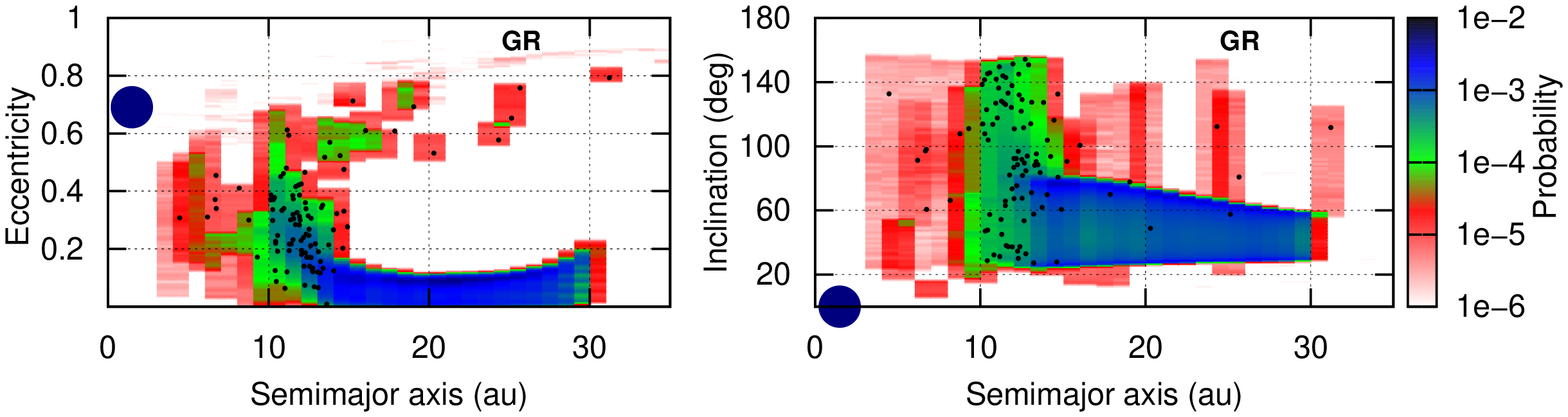}
\caption{Orbital planes ($a$, $e$) and ($a$, $i$). The color regions represent zones where the particles can be found with different levels of probability. The big blue circle illustrates the Jupiter-mass planet of the system and the black dots represent the surviving Type-F particles. Top panel: resulting system in absence of GR. Bottom panel: resulting system when the GR effects are included. The Jupiter-mass planet of the system has $a_{1} =$ 1.48 au and $e_{1}=$ 0.69.
}
\label{fig:fig3.6.2}
\end{figure*}

Now, we can use the integral of motion $f$ given by Eq.~\ref{eq:f-energia} to determine the extreme values of the inclination $i_{2}$ for libration trajectories including GR effects, which are associated to values of the ascending node longitude $\Omega_{2}$ of 90$^{\circ}$ or 270$^{\circ}$ as in the absence of GR. To do this, we must consider the conservation of $f$ on the separatrix between the minimum value of the ascending node longitude ($\Omega_{2} =$ 0$^{\circ}$, $i_{2} = i^{*}_{2}$) and the extreme values of the inclination ($\Omega_{2} =$ 90$^{\circ}$, $i_{2} = i^{\text{e}}_{2}$). It is worth noting that, when GR is included, the minimum of the ascending node longitude $\Omega_{2}$ on the separatrix is equal to 0$^{\circ}$ and is obtained for a value of the inclination $i_{2} = i^{*}_{2}$ given by Eq.~\ref{eq:imin}, while $i^{*}_{2} =$ 90$^{\circ}$ in absence of GR. From this, the extreme values of the inclination $i^{\text{e}}_{2}$ can be obtained from the following expression
\begin{eqnarray}
  \alpha \cos^{2} i^{\text{e}}_{2} + \beta \cos i^{\text{e}}_{2} + \gamma = 0,
  \label{eq:eq13}
\end{eqnarray}
where 
\begin{eqnarray}
 \alpha &=& 1 + 4 e^{2}_{1}, \\
 \label{eq:eq14}
 \beta &=& -A \frac{(1 - e^{2}_{2})^{2}}{(1 - e^{2}_{1})} \frac{a^{7/2}_{2}}{a^{9/2}_{1}},   \\
 \label{eq:eq15}
 \gamma &=& \frac{\beta^{2}}{4(1 - e^{2}_{1})} - 5e^{2}_{1},
 \label{eq:eq16}
\end{eqnarray}
where A is the constant given by Eq.~\ref{eq:constante-A}.

Figure~\ref{fig:fig4} shows the maximum and minimum inclination allowed for libration trajectories as a function of the planet's eccentricity $e_{1}$ for different values of the semimajor axis $a_{2}$ and the eccentricity $e_{2}$ of the test particle. In both panels, we set a value of 1 au for the planet's semimajor axis $a_{1}$. The extreme values of the inclination in absence of GR are illustrated by the black curves and they only depend on the planet's eccentricity, such as is indicated by Eq.~\ref{eq:imin-sinGR}. When the GR is included, the minimum and maximum values of the inclination depend on the semimajor axis $a_{1}$, eccentricity $e_{1}$, and mass of the planet $m_{1}$, as well as the semimajor axis $a_{2}$ and the eccentricity $e_{2}$ of the test particle, such is observed in Eqs.~(\ref{eq:eq13}-\ref{eq:eq16}). In general terms, our results indicate that the GR increases the minimum and maximum values of the inclination $i_{2}$ allowed for libration trajectories in comparison with those that correspond in absence of GR. It is worth noting that the smaller the particle's eccentricity, the higher the minimum and maximum values of the particle's inclination allowed for libration trajectories. This can be observed in both panels of Fig.~\ref{fig:fig4}, where the blue, green, and red curves illustrate the extreme values of the inclination of a Type-F particle for eccentricities $e_{2}$ of 0.1, 0.5, and 0.8, respectively, when the GR effects are included. Moreover, we infer that the greater the particle's semimajor axis $a_{2}$, the higher the minimum and maximum values of the particle's inclination. This can be seen in Fig.~\ref{fig:fig4}, where we use values of 15 au and 25 au for the particle's semimajor axis $a_{2}$ in the top and bottom panels, respectively.

As we will see in the next sections, an increase of the minimum and maximum inclinations allowed for libration trajectories has important consequences in the suppression and generation of orbital flips when the GR effects are included in the simulations. 

\subsection{Suppression of Type-F particles}

\citet{Naoz2017} suggested that the inclination of a given outer test particle is suppressed if its quadrupole precession is slower than the GR precession of the inner perturber of the system. Nevertheless, when GR precession acts on timescales larger (or even similar) than the quadrupolar precession, it can produce excitations in the particle's inclination. A behavior like this was previously observed by \citet{Naoz2013}, who studied the resonant post-Newtonian excitations of the eccentricity of an inner binary that is orbited by an outer massive perturber. In this context, similar results were derived by \citet{Fabrycky2007} and \citet{Liu2015}.

Figure~\ref{fig:fig3.2.1} shows the evolution in time of the inclination of three different particles of the same system with and without GR effects. The orbital parameters of the inner Jupiter-mass planet are $a_{1}=$ 1.32 au and $e_{1}=$ 0.56, while the GR precession timescale given by Eq.~\ref{eq:GR-timescale} is $t_{\text{GR}} =$ 130 Myr. In the left panel it can be seen that the suppression of the flips is efficient for a test particle whose libration period of inclination is only a small fraction of the GR precession timescale associated to the inner planet's orbit. In the middle panel, the orbital flip is suppressed when the libration period of the particle's inclination is similar to the planet's GR timescale, while the right panel shows that the orbital flip is suppressed for a particle whose libration period of inclination is significantly greater than the planet's GR precession timescale. In according to this, our results suggest that the suppression of the orbital flips is not sensitive to the period of libration associated to the particle's inclination.
                                   
A relevant result obtained from our study indicates that the suppression of the orbital flips is due to an increase in the minimum inclination allowed for the Type-F particles when the GR effects are included in the simulations. Thus, an increase in such minimum values of the inclination leads to a libration region which is more constrained in comparison with that obtained in absence of GR. In other words, the GR effects reduce the range of prograde inclinations for the libration regime. Remember that the minimum inclination allowed for libration trajectories, which is derived from Eqs.~(\ref{eq:eq13}-\ref{eq:eq16}), depends on the semimajor axis $a_{1}$, the eccentricity $e_{1}$, and the mass $m_{1}$ of the planet, as well as the semimajor axis $a_{2}$ and the eccentricity $e_{2}$ of the particle. According to that discussed in the previous section and illustrated in Fig.~\ref{fig:fig4}, the range of prograde inclinations for libration trajectories is more constrained for particles of low eccentricity and large semimajor axis.

To understand the suppression mechanism in the global structure of the outer reservoirs, we carried out occupation maps in the ($a$,$e$) and ($a$,$i$) orbital planes of a given system whose inner Jupiter-mass planet has a semimajor axis $a_{1} =$ 1.48 au and an eccentricity $e_{1}=$ 0.69. These maps are shown in Fig.~\ref{fig:fig3.6.2}, where the top and bottom panels illustrate the occupation maps of the corresponding system, which were developed from the results derived by the No-GR and GR simulations, respectively. Each of these plots shows the normalized time fraction spent by the test particles in different regions of the ($a$,$e$) and ($a$,$i$) orbital planes during the whole integration time. The color code indicates the permanence time spent in each region, where blue and red represent the most and least visited zones, respectively. We also remark that these plots are maps of probability since the portion of time spent in each zone is normalized. Moreover, in each map, the filled blue circle illustrates the inner Jupiter-mass planet, while the small black dots represent the surviving Type-F particles at the end of every simulation.

\begin{figure}
  \centering
  \includegraphics[angle=270, width=0.46\textwidth]{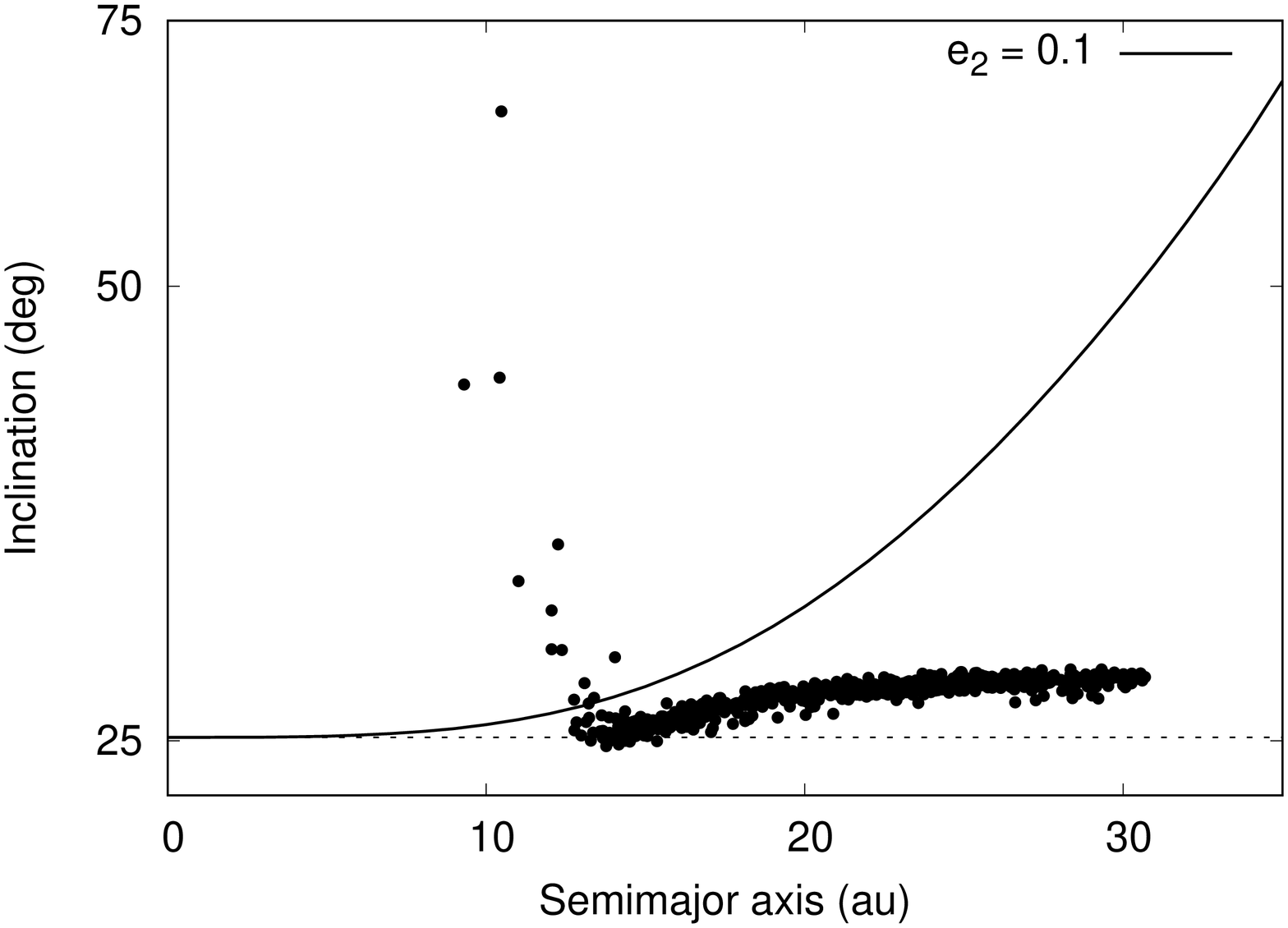} \\
  \includegraphics[angle=270, width=0.46\textwidth]{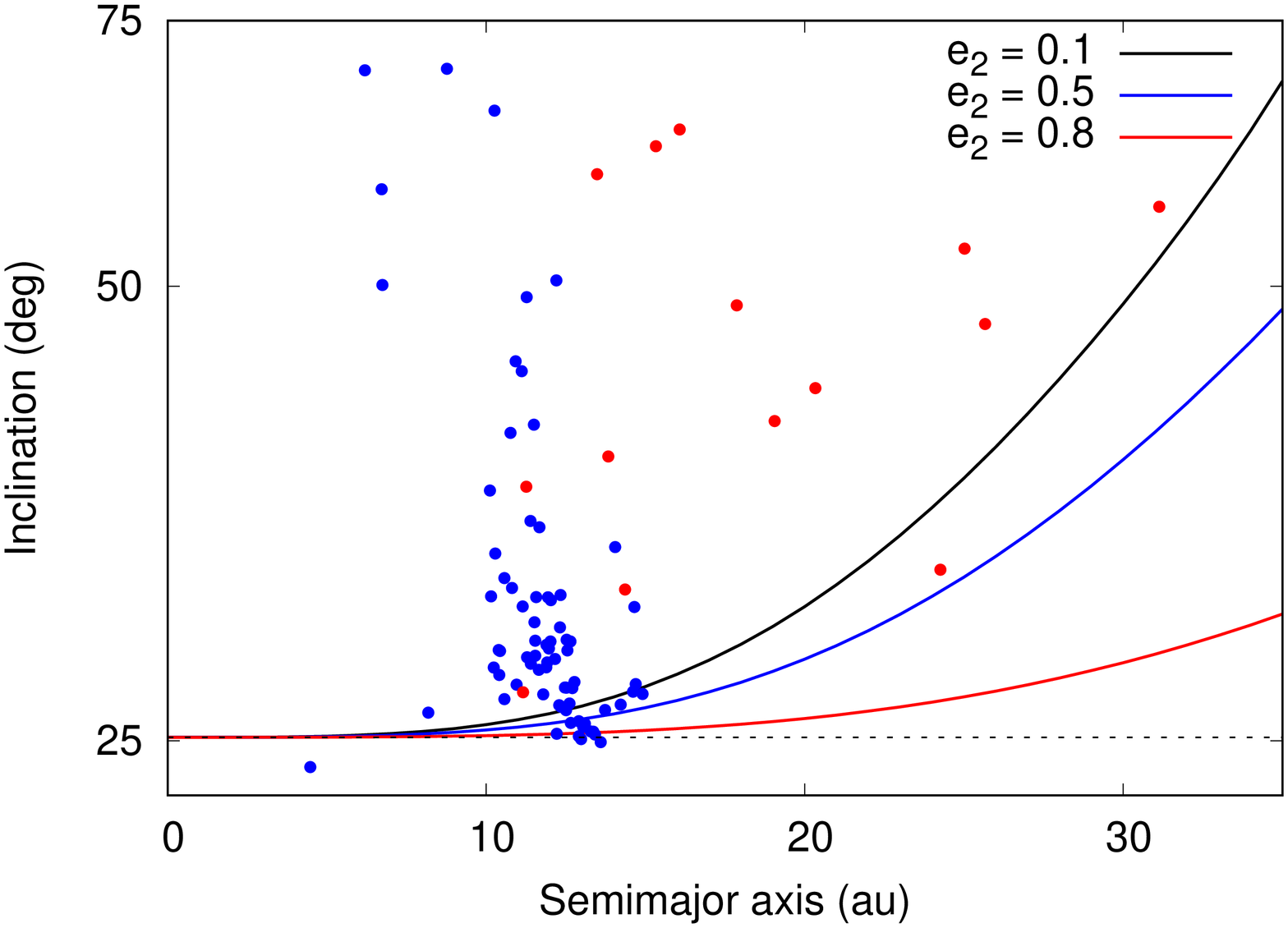}
\caption{Minimum inclination as a function of the semimajor axis of the Type-F particles resulting from the No-GR simulation corresponding to the system illustrated in the top panel of Fig.~\ref{fig:fig3.6.2}. The top panel shows the Type-F particles with eccentricities $e_{2} <$ 0.1 (filled black circles), while the bottom panel illustrates the Type-F particles with eccentricities $e_2$ between 0.1 and 0.5 (filled blue circles) and between 0.5 and 0.8 (filled red circles). The dashed horizontal black line represents the minimum inclination allowed for Type-F particles in absence of GR, which is equal to 25.1$^{\circ}$. The black, blue, and red curves represent the minimum inclination allowed for libration trajectories for values of the particle's eccentricity $e_{2}$ of 0.1, 0.5 and 0.8, respectively. The planet of the system has a semimajor axis and an eccentricity of $a_{1}=$ 1.48 au and $e_{1}=$ 0.69, respectively.
}
\label{fig:fig3.6.3}
\end{figure}

The top panel of Fig.~\ref{fig:fig3.6.2} shows that most of the surviving Type-F particles in the No-GR simulation are located between 10 au and 30 au and most of them have very low values of the eccentricity $e_{2}$ $\lesssim$ 0.1, but a few of them also survive with eccentricities $e_{2}$ between 0.1 and 0.9. However, when the GR effects are included, the dynamical properties of the outer particles significantly change, such as can be observed in the bottom panels of Fig.~\ref{fig:fig3.6.2}. In general terms, the orbital flips of particles with eccentricities $e_{2} \lesssim$ 0.1 are suppressed, while particles with eccentricities $e_{2} \gtrsim$ 0.1 keep their orbital flips showing a coupling between the inclination $i_{2}$ and the ascending node longitude $\Omega_{2}$. This change in the dynamical behavior of the particles is due to an increase in the minimum inclination allowed for libration trajectories when the GR effects are included. To understand it, we illustrate in Fig.~\ref{fig:fig3.6.3} the minimum inclination as a function of the semimajor axis of all Type-F particles that result from the No-GR simulation. In particular, the top panel shows the Type-F particles with eccentricities $e_{2} <$ 0.1 as filled black circles, while the bottom panel illustrates the Type-F particles with eccentricities $e_{2}$ between 0.1 and 0.5 (filled blue circles) and between 0.5 and 0.8 (filled red circles). To analyze the suppression mechanism of orbital flips, we must take into account the minimum inclination allowed for libration trajectories with and without GR effects. On the one hand, the dashed horizontal black line in both panels of Fig.~\ref{fig:fig3.6.3} illustrates the minimum inclination allowed for Type-F particles in absence of GR, which is given by Eq.~\ref{eq:imin-sinGR} and for this planetary system whose inner planet's eccentricity is $e_{1}=$ 0.69, has a constant value of 25.1$^{\circ}$. On the other hand, the black, blue, and red curves represent the minimum inclination allowed for Type-F particles for values of particle's eccentricity $e_{2}$ of 0.1, 0.5, and 0.8, respectively, when the GR effects are included. As the reader can see, the top panel of Fig.~\ref{fig:fig3.6.3} shows that most of the Type-F particles with eccentricities $e_{2} \lesssim$ 0.1 have minimum inclinations lower than the minimum inclination allowed for libration trajectories when the GR is included. Thus, the suppression mechanism of orbital flips is efficient over such particles of low eccentricity, which can be confirmed from the absence of Type-F particles with $e_{2} \lesssim$ 0.1 in the maps with GR of the bottom panels of Fig.~\ref{fig:fig3.6.2}. It is worth noting that a few Type-F particles of very low eccentricity and semimajor axes $a_{2} \lesssim$ 13 au are located above the black curve in the top panel of Fig.~\ref{fig:fig3.6.3}, which indicates that they should keep their orbital flips in the GR simulation. In fact, such Type-F particles survive in the maps with GR of Fig.~\ref{fig:fig3.6.2}. The Type-F particles that survive in the No-GR simulation with eccentricities $e_{2} \gtrsim$ 0.1 are illustrated in the bottom panel of Fig.~\ref{fig:fig3.6.3}. From this, it is possible to observe that most of them are located above the black, blue, and red curves, for which their minimum inclinations result to be higher than the minimum inclination allowed for libration trajectories when the GR is included. Thus, such particles should keep their orbital flips, which is confirmed from the existence of Type-F particles with eccentricities $e_{2}$ between 0.1 and 0.8 in the maps with GR of Fig.~\ref{fig:fig3.6.2}.

\begin{figure*}
\centering
\includegraphics[angle=270, width=0.99\textwidth]{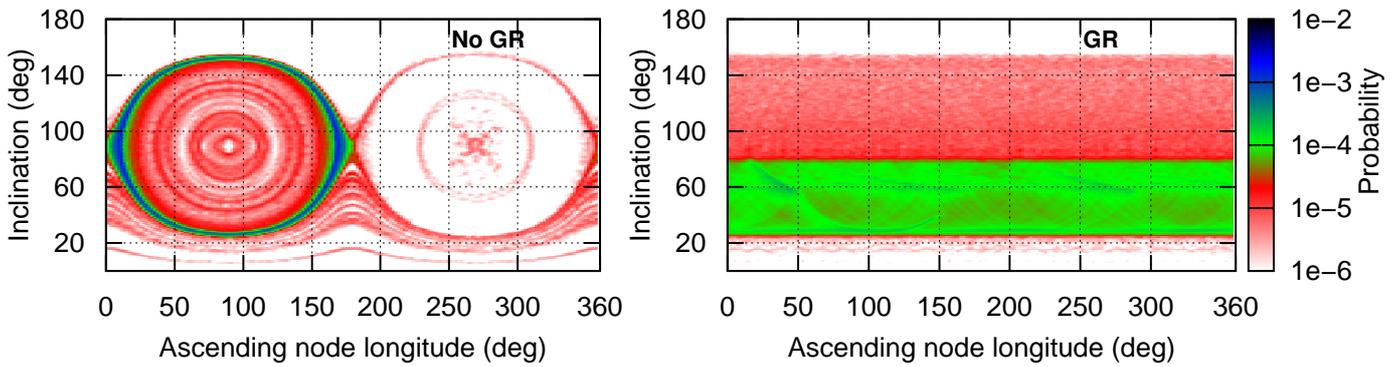}
\caption{Orbital planes ($\Omega_2$, $i_2$). The color regions represent zones where the particles can be found with different levels of probability.
Left panel: resulting system in absence of GR. Right panel: resulting system when the GR effects are included. The Jupiter-mass planet of the system has $a_{1} =$ 1.48 au and $e_{1}=$ 0.69.
}
\label{fig:fig3.6.2bis}
\end{figure*}

The suppression mechanism of orbital flips has been very efficient in the set of N-body simulations developed in the present work. In fact, in 9 of the 12 ``Type-F simulations'', the GR effects suppress between 5\% - 86\% of the Type-F particles of the No-GR simulations, with a median value for the suppression percentage of $\sim$ 40\%. In this context, the most relevant case corresponds to the system illustrated in Fig.~\ref{fig:fig3.6.2}, in which 86\% of the Type-F particles of the No-GR simulation are suppressed by the GR.

A natural consequence of the suppression of Type-F particles by GR effects is the increase in the number of the Type-P particles. In fact, an increase in the minimum inclination allowed for libration trajectories with GR leads to that particles on prograde initial orbits evolve following circulatory trajectories with prograde inclinations when the suppression mechanism of orbital flips is produced. In particular, in more than half of our simulations we obtained an increase of Type-P particles when the GR effects are included. In fact, the number of Type-P particles has been increased between 25\% - 91\% in the GR simulations in comparison with the number obtained from the No-GR simulations. It is worth noting that a case of special interest is that system illustrated in the maps of Fig.~\ref{fig:fig3.6.2}. Indeed, the number of Type-P particles in the resulting system from the No-GR and GR simulations is equal to 16 and 638, respectively. In this particular case, the generation of Type-P particles by the suppression of Type-F particles due to GR effects is very important, leading to significant changes in the dynamical structure of the outer reservoirs. Such behavior can be observed in the occupation maps ($\Omega_2$, $i_2$), which are represented in Fig.~\ref{fig:fig3.6.2bis}. Both panels show the differences in the ($\Omega_2$, $i_2$) plane of the same system illustrated in Fig.~\ref{fig:fig3.6.2} for the No-GR and GR simulations. On the one hand, the left panel illustrates the coupling between the inclination $i_2$ and the ascending node longitude $\Omega_2$ of the Type-F particles when the GR effects are absent in the N-body simulations. These correlations are represented for the blue and green libration regions, which are the zones most visited by the surviving Type-F particles. On the other hand, the right panel shows a significant increase of the number of Type-P particles in the system because of the suppression mechanism of Type-F particles turns out to be very efficient when the GR effects are included in the simulations. The effects of this suppression mechanism can be seen on the green regions in the right panel of  Fig.~\ref{fig:fig3.6.2bis}, which represent the zones most visited by the surviving Type-P particles.

\begin{figure*}
\centering
\includegraphics[angle=270, width=0.98 \textwidth]{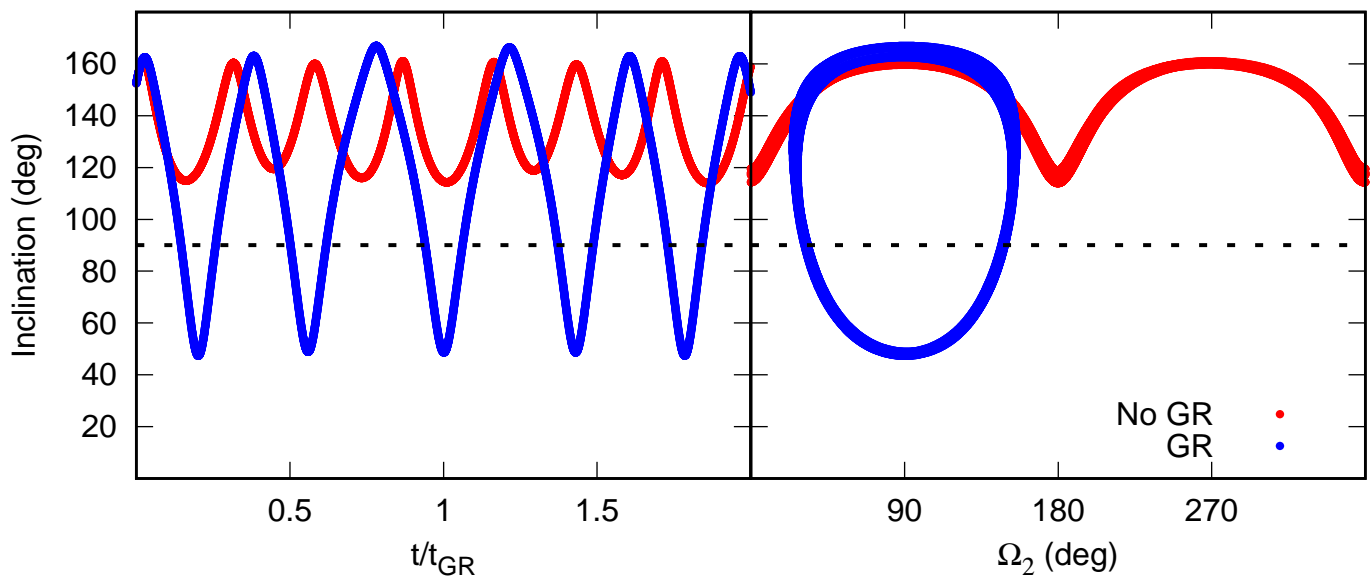}
\caption{Evolution of an outer particle that results from the No-GR (red curve) and GR (blue curve) simulations. The semimajor axis and eccentricity of the inner planet are $a_{1}=$ 0.95 au and $e_{1}=$ 0.73, and the orbit GR precession timescale is $t_{\text{GR}} =$ 39 Myr. The horizontal dashed black line represents a value of the inclination $i_{2}$ of 90$^{\circ}$. Left panel: evolution in time (normalized to $t_{\text{GR}}$) of the particle's inclination $i_2$. Right panel: evolution of the particle in the inclination $i_{2}$ vs. ascending node longitude $\Omega_{2}$ plane. The initial conditions of the particle are: $a_{2}=$ 35.75 au, $e_{2}=$ 0.85, $i_{2}=$ 152.48$\degr$, $\omega_{2}=$ 202.18$\degr$ and $\Omega_{2}=$ 41.03$\degr$.
}
\label{fig:fig3.3.1}
\end{figure*}

We conclude that the efficiency of the suppression mechanism of orbital flips when the GR effects are included in a given system is strongly dependent of the semimajor axes, eccentricities, and inclinations of the particles.

\subsection{Generation of Type-F particles}

A clear consequence of the increase in the maximum inclination allowed for libration trajectories with GR is the generation of Type-F particles from particles that evolve on retrograde orbits in absence of GR.    

Figure~\ref{fig:fig3.3.1} shows the evolution of a given particle that results from the No-GR (red curve) and GR (blue curve) simulations, in a system whose inner Jupiter-mass planet has a semimajor axis $a_{1}$ of 0.95 au and an eccentricity $e_{1}$ of 0.73, while the GR precession timescale given by Eq.~\ref{eq:GR-timescale} is of 39 Myr. On the one hand, the left panel of Fig.~\ref{fig:fig3.3.1} illustrates the time evolution of the particle's inclination $i_2$. On the other hand, the evolution of the inclination $i_2$ and the ascending node longitude $\Omega_2$ of the test particle is shown in the right panel of Fig.~\ref{fig:fig3.3.1}. From this, the test particle evolves on retrograde orbits in absence of GR, while when the GR effects are included in the simulations, the orbit of such a particle flips from retrograde to prograde and back again showing a coupling between the inclination $i_2$ and the ascending node longitude $\Omega_2$. The GR effects increase the maximum inclination allowed for Type-F particles, which enlarges the range of retrograde inclinations for libration trajectories.

The generation mechanism of orbital flips from particles that evolve on retrograde orbits in absence of GR is present in half of all our simulations. In fact, in 6 of the 12 ``Type-F simulations'', 3.7\% - 31.7\% of Type-R particles of the No-GR simulations becomes Type-F particles when the GR effects are included. This generation percentage has a median value of $\sim$ 11.6\%.

It is important to remark that the number of Type-F particles produced from particles on retrograde orbits in absence of GR is lower than the number of Type-P particles generated from the suppression of Type-F particles discussed in the previous subsection. It is worth noting that this result is a clear consequence of the initial conditions used in the present study. Remember that we assumed as initial conditions of our work the orbital parameters of the 12 ``Type-F simulations'' defined from the Paper 1 immediately after the instability event, when a single Jupiter-mass planet survives in the system. In fact, Fig.~\ref{fig:fig2}, which reflects the initial orbital parameters of those 12 ``Type-F simulations'' used in the present work, shows that the number of particles with initial inclinations higher than 90$\degr$ is much lower than the number associated with inclinations less than 90$\degr$. Thus, we conclude that, when the GR is included in the simulations, the mechanism of suppression of orbital flips and generation of Type-P particles is more efficient than the production of Type-F particles from particles that evolve on retrograde orbits in absence of GR.

\subsection{Flipping particles without coupling between inclination and ascending node longitude}

The GR simulations show the existence of particles whose orbital planes flip reaching prograde and retrograde inclinations without experiencing a coupling between the inclination $i_{2}$ and the ascending node longitude $\Omega_{2}$. Due the absence of coupling between $i_{2}$ and $\Omega_{2}$, they are not Type-F particles and so, we simply call them ``flipping particles''. 

A flipping particle does not show coupling between the orbital inclination $i_{2}$ and the ascending node longitude $\Omega_{2}$ because $i_{2}$ is always less than the necessary value to obtain extreme values of $\Omega_{2}$ in any point of the trajectory, which is given by Eq.~\ref{eq:imin}. Thus, $\Omega_{2}$ is not constrained between two specific values for which such an angle evolves in a circulatory regime.

\begin{figure}
\centering
\includegraphics[angle=270, width=0.48 \textwidth]{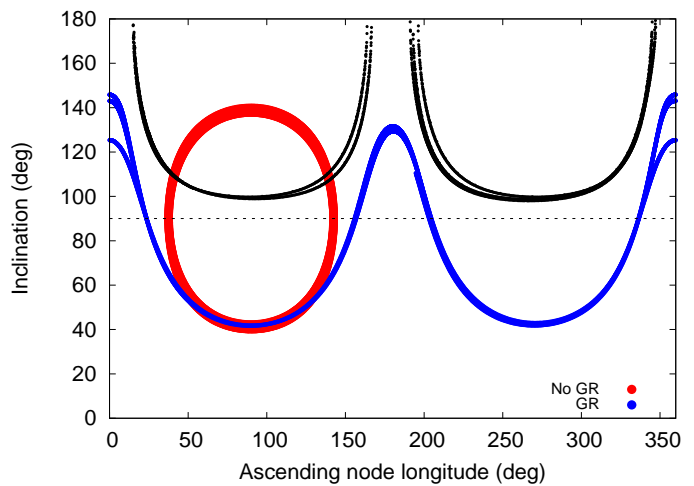}
\caption{Evolution of the inclination $i_{2}$ and ascending node longitude $\Omega_{2}$ of a given particle resulting from the No-GR (red curve) and GR simulations (blue curve). The black dots illustrate the values of the particle's inclination $i_{2}$ for which the ascending node longitude reaches extreme values given by Eq.~\ref{eq:imin}. The horizontal dashed black line represents a value of the inclination $i_{2}$ of 90$^{\circ}$. The planet of the system has a semimajor axis $a_{1}$ of 0.867 au and an eccentricity $e_{1}$ of 0.8. The initial conditions of the particle are: $a_{2}=$ 20.792 au, $e_{2}=$ 0.705, $i_{2}=$ 51.09$\degr$, $\omega_{2}=$ 157.95$\degr$ and $\Omega_{2}=$ 52.9$\degr$.  
}
\label{fig:flipingparticles}
\end{figure}

This behavior can be observed in Fig.~\ref{fig:flipingparticles}, which shows the evolution of the inclination $i_{2}$ and the ascending node longitude $\Omega_{2}$ of a test particle of a given system in No-GR (red curve) and GR (blue curve) simulations, which has an initial semimajor axis $a_{2} =$ 20.792 au and an high initial eccentricity $e_{2} =$ 0.705. In such a system, the inner planet has a semimajor axis $a_{1} =$ 0.867 au and an eccentricity $e_{1} =$ 0.8. In the No-GR simulation, such a particle flips experiencing a coupling between $i_{2}$ and $\Omega_{2}$. In the GR simulation, the particle evolves on a circulation trajectory since the inclination $i_{2}$ at $\Omega_{2} =$ 90$^{\circ}$ and 270$^{\circ}$ is higher than the inclination $i_{2}$ allowed for libration trajectories, which is given by Eq.~\ref{eq:eq13}. Beyond this, the particle's orbit flips from prograde to retrograde and back again reaching a significant maximum inclination $i_{2} =$ 145.825$^{\circ}$. In fact, the particle can reach retrograde inclinations without experiencing a coupling between $i_{2}$ and $\Omega_{2}$ since the inclination $i_{2}$ is always less than the necessary value to obtain extreme values of $\Omega_{2}$ in any point of the trajectory, which is calculated from Eq.~\ref{eq:imin} and it is illustrated by black points in Fig.~\ref{fig:flipingparticles}.

Our study indicates that 6 of 12 ``Type-F simulations'' that include GR effects do not produce flipping particles, while, in the others 6 GR simulations, the number of flipping particles ranges between 23\% - 77\% of the number of Type-F particles of the system.

\subsection{Type-R particles with coupling between inclination and ascending node longitude}

Our study shows that of the 161 Type-R particles that survived in all our GR simulations, only 4 of them experience a coupling between the inclination $i_2$ and the ascending node longitude $\Omega_2$.

\begin{figure}
\centering
\includegraphics[angle=270, width=0.48 \textwidth]{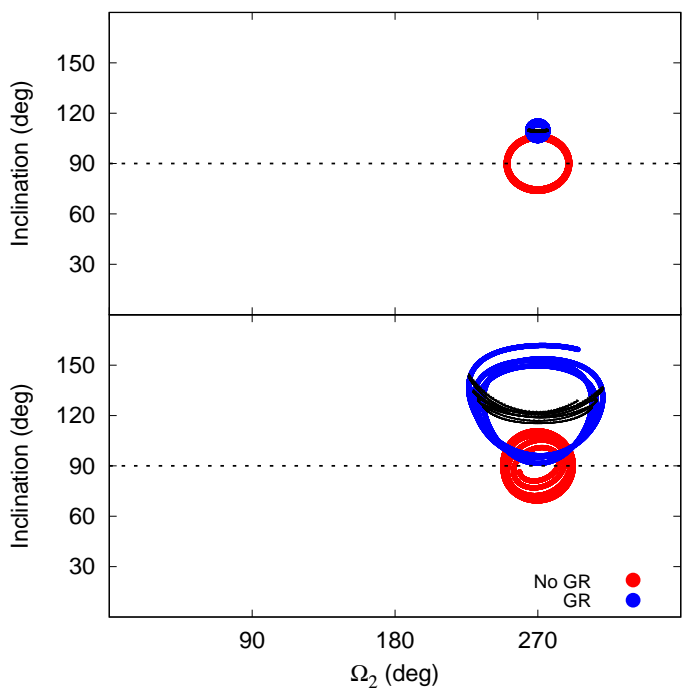}
\caption{Evolution of the inclination $i_2$ and ascending node longitude $\Omega_2$ of a given particle resulting from the No-GR (red curve) and GR simulations (blue curve). The black dots illustrate the values of the particle's inclination $i_2$ for which the ascending node longitude reaches extreme values, which is given by Eq.~\ref{eq:imin}. The horizontal dashed black line represents a value of the inclination $i_2$ of 90$^{\circ}$. The planet of the system has a semimajor axis $a_1$ of 0.736 au and an eccentricity $e_1$ of 0.529. Top panel: the initial conditions of the particle are: $a_{2}=$ 15.05 au, $e_{2}=$ 0.143, $i_{2}=$ 106.58$\degr$, $\omega_{2}=$ 93.67$\degr$ and $\Omega_{2}=$ 272$\degr$. Bottom panel: the initial conditions of the particle are: $a_{2}=$ 28.39 au, $e_{2}=$ 0.784, $i_{2}=$ 98.63$\degr$, $\omega_{2}=$ 90.51$\degr$ and $\Omega_{2}=$ 251.74$\degr$. 
}
\label{fig:retro-acopla}
\end{figure}

Figure~\ref{fig:retro-acopla} shows the evolution of the inclination $i_{2}$ and the ascending node longitude $\Omega_{2}$ of two different outer particles of a same system resulting from No-GR (red curve) and GR (blue curve) simulations. The inner planet associated to such a system has a semimajor axis $a_{1} =$ 0.736 au and an eccentricity $e_{1} =$ 0.529. In the No-GR simulations, the orbital plane of every particle flips showing a coupling between $i_{2}$ and $\Omega_{2}$, which is observed in the red curves illustrated on the two panels of Fig.~\ref{fig:retro-acopla}. When the GR effects are included in the simulations, the dynamical evolution of such particles shows significant changes. In the top panel, the blue curve describes a libration trajectory where the minimum inclination $i_{2}$ ranges between 103.7$^{\circ}$ and 106.25$^{\circ}$, and the values of $i_{2}$ to obtain extreme values of $\Omega_{2}$, which are illustrated as black points, slightly oscillate between 109$^{\circ}$ and 110.2$^{\circ}$. The ascending node longitude $\Omega_{2}$ is constrained between two specific values, while the particle only acquires retrograde inclinations during its whole evolution. In the bottom panel, the blue curve illustrates a libration trajectory where the minimum inclination $i_{2}$ ranges between 91.4$^{\circ}$ and 96.1$^{\circ}$, and the values of $i_{2}$ that produce extreme values of $\Omega_{2}$, which are represented as black points, significantly oscillate between 115.32$^{\circ}$ and 144.75$^{\circ}$. As the particle represented in the top panel, the inclination $i_2$ only adopts retrograde values while the ascending node longitude $\Omega_2$ maintains constrained between two given values.

As we have already mentioned, the generation of this class of particles is not efficient in the GR simulations developed in the present research. However, we consider that it is very important to carry out a detailed description about 
them in order to reach a correct understanding of the structure dynamical of the outer small body reservoirs.

\section{Conclusions}

In this paper, we have studied the role of GR on the evolution of icy small body reservoirs under the effects of an eccentric inner Jupiter-mass planet produced by a planetary scattering event. To carry out our research, we have selected 12 scenarios of work of the Paper 1 adopting as initial conditions of the test particles the orbital parameters immediately after of the dynamical instability event when a single Jupiter-mass planet survives in each system. To study how the GR effects modify the orbital properties of the test particles, we have carried out two sets of N-body simulations, one of which only included gravitational forces in the evolution ({\it No-GR simulations}), while the other one also incorporated GR effects in the integration of each of such systems of work ({\it GR simulations}).

In absence of GR, the ascending node longitude $\Omega_2$ of a particle can evolve in two different regimes: libration or circulation. In the libration regime, the ascending node longitude $\Omega_2$ adopts extreme values for $i_2 =$ 90$\degr$, while the minimum and maximum values of the particle's inclination allowed for libration trajectories only depends on the eccentricity $e_1$ of the inner planet. When the GR effects are included in the simulations, the properties of the particles that evolve on libration mode significantly change. In fact, the extreme values of the ascending node longitude $\Omega_2$ are obtained for retrograde inclinations, while the minimum and maximum inclination allowed for libration trajectories depends on the mass $m_1$, semimajor axis $a_1$ and eccentricity $e_1$ of the inner planet, as well as the semimajor axis $a_2$ and eccentricity $e_2$ of the particle. In a given system, the smaller the eccentricity and the greater the semimajor axis of the particle, the higher the minimum and maximum values of the particle's inclination allowed for libration trajectories. In this context, our results suggest that the minimum and maximum inclination allowed for the libration regime increase in comparison with those values derived without GR effects. According to this, if the GR is included in the simulations, the range of prograde (retrograde) inclinations of the libration region is reduced (increased) respect to that obtained in absence of GR. It is important to remark that we have derived analytical expressions up to quadrupole level of the secular approximation for the inclination that produces the extreme values of the ascending node longitude (Eq.~\ref{eq:imin}), as well as for the minimum and maximum inclination allowed for libration trajectories (Eqs.~\ref{eq:eq13}-\ref{eq:eq16}) when the GR effects are included.  

These points have important consequences in the dynamical evolution of the outer particles of a given system. On the one hand, if the range of prograde inclinations for libration trajectories is reduced, the GR results to be an efficient suppression mechanism of Type-F particles, which leads to generation of Type-P particles. According to that discussed in the previous paragraph, notice that the suppression of Type-F particles is an extremely efficient mechanism for very low eccentricity orbits. On the other hand, if the GR enlarges the range of retrograde inclinations for the libration region, Type-R particles in No-GR simulations can evolve as Type-F particle when the GR effects are included. However, the suppression mechanism of Type-F particles is the most important one in our simulations due to a strong preponderance of prograde orbits in the initial conditions of the test particles immediately after the dynamical instability event modeled in Paper 1.

We remark that a general analysis concerning the structure of the outer small body populations resulting from the GR simulations shows a great diversity of dynamical behaviors. In fact, besides the Type-P, -R and -F particles that also are present in the outer reservoirs resulting from the No-GR simulations, we distinguished two new classes of particles when the GR effects are included. First, the so-called flipping particles, which flip from prograde to retrograde and back again along their evolution without a coupling between the inclination $i_2$ and the ascending node longitude $\Omega_2$. Then, particles that evolve on retrograde orbits showing a strong coupling between the inclination $i_2$ and the ascending node longitude $\Omega_2$. 

The present research allows us to infer that the GR effects can modify the dynamical structure of outer reservoirs formed from planetary scattering events and evolve under the influence of an inner and eccentric massive perturber. These results have important consequences in the evolution of debris disks associated to planetary systems that host a single inner gaseous giant on an eccentric orbit. In fact, the GR effects must be included in the models aimed at analyzing the debris disks in that class of systems in order to carry out a correct description concerning the dynamic of such structures. This task is crucial for future studies that develop a collisional evolution of the debris disks with the main goal to quantify their production rate of dust. A comparative analysis between our numerical results and observational data obtained from {\it Spitzer} will allow us to constraint our theoretical models and to strengthen our understanding about the collisional and dynamical evolution of the outer reservoirs associated to planetary systems that host an inner and eccentric gaseous giant.

\begin{acknowledgements}
This work was partially financed by CONICET and Agencia de Promoci\'on Cient\'{\i}fica, through the PIP 0436/13 and PICT 2014-1292.
MZ, GdE and RD acknowledge the financial support by FCAGLP for extensive use of its computing facilities. SN acknowledges partial support from a Sloan Foundation Fellowship. Finally, we thank the anonymous referee for valuable suggestions, which helped us to improve the manuscript.
\end{acknowledgements}
\bibliographystyle{aa}      
\bibliography{zanardi_GR}   

\end{document}